\documentclass{emulateapj}
\newcommand{\be}{\begin{equation}}
\newcommand{\ee}{\end{equation}}
\newcommand{\bea}{\begin{eqnarray}}
\newcommand{\eea}{\end{eqnarray}}

\begin{document}

\title{Cosmic Ray Propagation: Nonlinear Diffusion Parallel and Perpendicular to Mean Magnetic Field}
\author{Huirong Yan\altaffilmark{1} and A. Lazarian\altaffilmark{2}}

\altaffiltext{1}{Canadian Institute of Theoretical Astrophysics, 60 St. George Street, Toronto, ON M5S 3H8, Canada; yanhr@cita.utoronto.ca}
\altaffiltext{2}{Astronomy Department, University of Wisconsin, Madison, WI 53706; alazarian@wisc.edu}

\begin{abstract}
We consider the propagation of cosmic rays in turbulent magnetic fields. We use
the models of magnetohydrodynamic turbulence that were tested in numerical simulations, in which the turbulence is injected on large scale and cascades to small scales. Our attention is focused on the models of the strong turbulence, but we also briefly discuss the effects that the weak turbulence and the slab Alfv\'enic perturbations can have. The latter are likely to emerge as a result of instabilities with in the cosmic ray fluid itself, e.g., beaming and 
gyroresonance instabilities of cosmic rays. To describe the interaction of cosmic rays with magnetic perturbations we develop a non-linear formalism that extends the ordinary Quasi-Linear Theory (QLT) that is routinely used for the purpose. This allows us to avoid the usual problem of 90 degree scattering and enable our computation of the mean free path of cosmic rays. We apply the formalism to the cosmic ray propagation in the galactic halo and in the Warm Ionized medium (WIM). In addition, we address the issue of the transport of cosmic rays perpendicular to the mean magnetic field and show that the issue of cosmic ray subdiffusion (i.e., propagation with retracing the trajectories backwards, which slows down the diffusion) is only important for restricted cases when the ambient turbulence is far from what numerical simulations suggest to us. As a result, this work provides formalism that can be applied for calculating cosmic ray propagation in a wide variety of circumstances.     

\end{abstract}


\section{INTRODUCTION}

The propagation and acceleration of cosmic rays (CRs) is governed by their interactions with magnetic fields. Astrophysical magnetic fields are turbulent and, therefore, the resonant and non-resonant (e.g., transient time damping, or TTD) interaction of cosmic rays with MHD turbulence is the accepted principal mechanism to scatter and isotropize cosmic rays (see Schlickeiser 2002). In addition, efficient scattering is essential for the acceleration of cosmic rays. 
For instance, scattering of cosmic rays back into the shock is a
vital component of the first order Fermi acceleration (see Longair
1997). At the same time, stochastic acceleration by turbulence is 
entirely based on scattering.

It is generally accepted that properties of turbulence are vital for the
correct description of CR propagation. Historically, the most widely used
model is the model composed of slab perturbations and 2D MHD
perturbations (see Bieber, Smith, \& Matthaeus 1988). The advantage of
this empirical model is its simplicity and the ability to account
for the propagation of CRs in magnetosphere given a proper partition of the energy between the two types of modes.

Numerical simulations (see Cho \& Vishniac 2000,
Maron \& Goldreich
2001, M\"uller \& Biskamp 2000, Cho, Lazarian \& Vishniac 2002, 
Cho \& Lazarian 2002, 2003, see also book of Biskamp 2003, as well as,
Cho, Lazarian \& Vishniac (2003) and Elmegreen \& Scalo 2004 for reviews),
however, do not show 2D modes, but instead show Alfv\'enic modes that exhibit scale-dependent anisotropy consistent with predictions in Goldreich \& Sridhar (1995, henceforth GS95). 
The approach in the latter work makes productive use of the earlier advances in understanding of MHD turbulence, that can be traced back to Iroshnikov (1963) and Kraichnan (1965) work and the classical work that followed (see Montgometry \& Turner 1981, Higdon 1984, Montgomery, Brown \& Matthaeus 1987). 

A careful analysis shows that there is no big gap between the Reduced MHD and the GS95 model. In fact, it was shown in Lazarian \& Vishniac (1999) that the numerical results in
Matthaeus et al. (1998) are consistent with GS95 predictions. While particular aspects of the GS95 model, e.g., the particular value of the spectral index, are the subject of controversies\footnote{To address quantitatively these controversies, we need much better numerical resolution. For instance, hydrodynamic turbulence simulations in Kritsuk et al. (2007) showed that only starting
with the 1028$^3$ numerical cubes the bottleneck effects stop dominating the
measured spectral slope. While the simulations in Kowal \& Lazarian (2007)
show that the bottleneck is less important for their MHD code, the exact
value of the spectral slope is still uncertain. At the same time, the 
particular theoretically-predicted features of turbulence, for instance, the existence of the scale-dependent anisotropy, can be reliably established, while the exact scaling of this dependence, e.g., like $k_{\|}\sim k_{\bot}^{2/3}$ as in GS95 model or $k_{\|}\sim k_{\bot}^{\alpha*}$, where $\alpha*<2/3$ (see Beresnyak \& Lazarian 2006), also require higher resolution simulations.}  
(see M\"uller \& Biskamp 2000, Boldyrev 2005, 2006, Beresnyak \& Lazarian 2006, Gogoberidze 2006, Mason et al. 2007), we think that, at present, GS95 model provides a good starting for developing models of CR scattering.
as was done in Chandran (2000), Yan \& Lazarian (2002, 2004, henceforth YL02, Paper I, respectively), Brunetti \& Lazarian (2007) etc. In particular, the
latter three papers used the decomposition of MHD turbulence over Alfv\'en, slow and fast modes as in Cho \& Lazarian (2003) and identify the fast modes as the major source of CR scattering in interstellar and intracluster medium.

However, while the turbulence injected on large scales may correspond to GS95 model and its extensions to compressible medium (Lithwick \& Goldreich 2001, Cho \& Lazarian 2002, 2003), one should not disregard the possibilities of generation of additional perturbations by CR themselves. Indeed,  the slab Alfv\'enic perturbation can be created, e.g., via streaming instability (see Wentzel 1974, Cesarsky 1980) or kinetic gyroresonance instability (see its application for CR transport in Lazarian \& Beresnyak 2006). These perturbations, that are present for a range of CRs energies (e.g., $\lesssim 100$GeV 
for the instabilities above in ISM) owing to non-linear damping 
arising from ambient turbulence (YL02, Paper I, Farmer \& Goldreich 2004, Lazarian \& Beresnyak 2006), should also be incorporated into the comprehensive models of CR propagation and acceleration.

At present, the propagation of the CRs is an advanced theory, which makes use both of analytical studies and numerical simulations. However, these advances have been done within the turbulence paradigm which
is being changed by the current research in the field.
As we discussed above, instead of the empirical 2D+slab model of turbulence, numerical simulations suggest anisotropic Alfv\'enic modes (an analog of 2D, but not an exact one, as the anisotropy changes with the scale involved) + fast modes
or/and slab modes. This calls for important revisions of the CR propagation, which is the subject of the current paper.

The perturbations of turbulent magnetic field are usually accounted for by direct numerical scattering simulations (Giacalone \& Jokipii 1999, Mace et al 2000, Qin at al. 2002) or by quasi-linear theory, QLT (see Jokipii 1966, Schlickeiser 2002). The problem with direct numerical simulations of scattering is that the present-day
MHD simulations have rather limited inertial range. At the same time, creating synthetic turbulence data which would correspond to scale-dependent anisotropy in respect to the local magnetic field (which corresponds, e.g., to GS95 model) is challenging and has not been practically realized, as far as we know. 

While QLT allows easily to treat the CR dynamics in a local magnetic
field system of reference, a key assumption in QLT, that the particle's orbit is unperturbed, makes one wonder about the limitations of the approximation. Indeed, while QLT provides simple physical insights into scattering, it is known to have problems. For instance, it fails in treating $90^o$ scattering  (see Jones, Birmingham \& Kaiser 1973, 1978; V\"olk 1973, 1975; Owens 1974;
 Goldstein 1976; Felice \& Kulsrud 2001) and perpendicular transport
 (see K\'ota \& Jokipii 2000, Matthaeus et al. 2003).

Indeed, many attempts have been made to improve the QLT and various non-linear
 theories have been attempted (see Dupree 1966, V\"olk 1973, 1975, 
Jones, Kaiser 
\& Birmingham 1973, Goldstein 1976). Currently we observe a surge
of interest in finding way to go beyond QLT. Those include recently developed nonlinear guiding center theory (see Matthaeus et al. 2003), weakly nonlinear theory (Shalchi et al. 2004), second-order 
quasilinear theory (Shalchi 2005a) (see also Shalchi 2006, Webb et al. 2006, Qin 2007, Le Roux \& Webb 2007). At the same time, most of the analysis so far has been confined to traditional 2D+slab models of MHD turbulence. 
Following the reasoning above, we think that it is important to extend the work to the non-linear treatment of CR scattering to models MHD turbulence that are supported by numerical simulations.

Propagation of CRs perpendicular to the mean magnetic field is another important problem in which QLT encounters serious difficulties.
Compound diffusion, resulting from the convolution of diffusion along the magnetic field line and diffusion of field line perpendicular to mean field direction, has been invoked to discuss transport of cosmic rays in the Milky Way (Getmantsev 1963; Lingenfelter, Ramary \& Fisk 1971; Allan 1972). The role of compound diffusion in the acceleration
 of CRs at quasi-perpendicular shocks were investigated by Duffy et al. (1995) and Kirk et al. (1996).  

Indeed, the idea of CR transport in the direction perpendicular to the mean magnetic field being dominated by the field line random walk 
(FLRW, Jokipii 1966, 
Jokipii \& Parker 1969, Forman et al. 1974) can be easily justified
only in a restricted situation where the turbulence perturbations are small and CRs do not scatter backwards to retrace their trajectories. If the latter is not true, the particle motions are subdiffusive, 
i.e., the squared distance diffused growing as not as $t$ but as $t^{\alpha}$, $\alpha<1$, e.g., $\alpha=1/2$ (K\'ota \& Jokipii 2000, Mace et al 2000, Qin at al. 2002, Shalchi 2005b).
If true, this could indicate a substantial shift in the paradigm of CR transport, a shift that surely dwarfs a modification of magnetic turbulence model from the 2D+slab to a more simulation-motivated model that we deal here.

It was also proposed that with substantial transverse structure, {\it i.e.}, transverse displacement of field lines, perpendicular diffusion is recovered (Qin et al 2002). Is it the case of the MHD turbulence models we deal with?

How realistic is the subdiffusion in the presence of turbulence? The answer for this question apparently depends on the models of turbulence chosen. In this paper we again seek the answer for this question within domain of numerically tested models of MHD turbulence.

There are three major thrusts of the paper:\\ 
I. Extend QLT by taking into account magnetic mirroring effect on large scales.\\
II. Describe CR propagation in Milky Way (e.g., calculate CR mean free path for different phases of ISM).\\
III. Address the problem of perpendicular transport of CR.

In what follows, we discuss the cosmic ray transport in incompressible turbulence in \S2. We shall describe the \S2.1 dispersion of guiding center of CRs and introduce the broadened resonance function to replace the $\delta$ function in QLT, following which we shall discuss the scattering in strong and weak incompressible turbulence respectively in \S2.2 and \S2.3. Then we shall consider the scattering by fast modes in \S3 and apply the analysis to ISM and get mean free path for different phases of ISM (\S4). In \S5, we shall study the perpendicular transport of cosmic rays on both large and small scales. We shall also discuss the applicability of the subdiffusion. Discussion and summary are provided in \S6 and \S7 respectively.

\section{CR Transport in incompressible turbulence}

\subsection{General formalism}

It was demonstrated that scattering by the Alfv\'enic turbulence is substantially suppressed due to its anisotropy (Chandran 2000, YL02). On the other hand, resonant mirror interaction (so-called transit time damping or TTD) can arise from the slow modes (also known as the pseudo Alfv\'en modes in incompressible limit) and it is not subjected to the suppression from anisotropy. One may speculate that TTD is the alternative in this case.

The only requirement for TTD is that the projected particle speed is comparable to phase speed of the magnetic field compression\footnote{ The compressions of the magnetic field in incompressible turbulence are related the pseudo-Alfv\'en mode, which is the limiting case of the slow mode (see Alfv\'en \& F\"althammar 1963).} ($v_\parallel\simeq \omega/k_\parallel$) for particles to have enough collisions with the moving magnetic mirrors before they leak out of them. In QLT, this means that only particles with a specific pitch angle 
can be scattered. For the rest of the pitch angles, the interaction is still dominated by gyroresonance, which efficiency is negligibly small for the Alfv\'enic anisotropic turbulence. The quantitative treatment of this interaction is in YL02, which used the empirical magnetic field fluctuation tensor from Cho et al. (2002), provided the reduction factor of twenty orders of magnitude for CR of 100 GeV energies. This leads to very large mean free path of CRs. With the resonance broadening, however, we expect that wider range of pitch angle can be scattered through TTD, including $90^o$. 

The basic assumption of the quasi-linear theory is that particles follow unperturbed orbits. In reality, particle's pitch angle varies gradually with the variation of the magnetic field due to conservation of adiabatic invariant $v_\bot^2/B$, where $B$ is the total strength of the magnetic field (see Landau \& Lifshits 1975). Since B is varying in turbulent field, so is the projection of the particle speed $v_\bot$ and $v_\|$.
 This results in broadening of the resonance. Indeed, the average uncertainty of parallel speed $\Delta v_\parallel$ is given by (see V\"olk 1975)
\bea
\frac{\Delta v_\parallel}{v_\perp}&=&\frac{<(B-B_0)^2>^{1/4}}{B_0^{1/2}}\simeq\left[\frac{<\delta B_\parallel^2>}{B_0^2}+o\left(\frac{<(\delta B_\bot)^2>^2}{B_0^4}\right)\right]^{1/4}
\label{deltav}
\eea
The variation of the velocity is mainly caused by the magnetic perturbation $\delta B_\|$ in the parallel direction. This is true even for the incompressible turbulence we discussion in this section. For the incompressible turbulence, the parallel perturbation arises from the pseudo-Alfv\'en modes. The perpendicular perturbation $\delta B_\bot$ is higher order effect, which we shall neglect in this paper.

The propagation of a CR can be described as a combination of a motion of its guiding center and CR's motion about its guiding center. 
Because of the dispersion of the pitch angle $\Delta\mu$ and therefore of the parallel speed $\Delta v_\|$, the guiding center is perturbed about the mean position $<z>=v\mu t$ as they move along the field lines. As a result, the perturbation $\delta B({\bf x},t)$  that the CRs view when moving along the field gets a different time dependence. The characteristic phase function $e^{ik_\|z(t)}$ of the perturbation $\delta B({\bf x},t)$ deviates from that for plane waves. Assuming the guiding center has a Gaussian distribution along the field line, one obtains
\be
f(z)=\frac{1}{\sqrt{2\pi}\sigma_z}e^{-\frac{(z-<z>)^2}{2\sigma_z^2}},
\label{gauss}
\ee
Integrating over z, one gets 
\be
\int_{-\infty}^{\infty} dze^{ik_\| z}f(z)= e^{ik_\|<z>}e^{-k_\|^2\sigma_z^2/2}. 
\label{phase}
\ee
From Eq.(\ref{deltav}), we obtain 
\be
\sigma_z^2=<\Delta v_\|^2>t^2=v_\bot^2\left(\frac{<\delta B_\parallel^2>}{B_0^2}\right)^\frac{1}{2}t^2.
\ee

Perturbation $\delta B_\|$ exists owing to the pseudo-Alfv\'en  modes in the incompressible turbulence. Insert the Eq.(\ref{phase}) into the expression of $D_{\mu\mu}$ (see V\"olk 1975, Paper I), we obtain

\bea
D_{\mu\mu}&=&\frac{\Omega^2(1-\mu^2)}{B_0^2}\int d^3k\sum_{n=0}^{\infty}R_n(k_{\parallel}v_{\parallel}-\omega\pm n\Omega)\nonumber\\
&&\left[I^A({\bf k})\frac{n^2J_n^2(w)}{w^2}+\frac{k_\|^2}{k^2}J^{'2}_n(w)I^M({\bf k})\right],
\label{general}
\eea 
Following are the definitions of the parameters in the above equation. $\Omega, \mu$ are the Larmor frequency and pitch angle cosine of the CRs. $J_n$ represents Bessel function, and $w=k_\bot v_\bot/\Omega=k_\bot LR\sqrt{1-\mu^2}$, where $R=v/(\Omega l)$ is the dimensionless rigidity of the CRs, $L$ is the injection scale of the turbulence. $k_\bot, k_\|$ are the components of the wave vector ${\bf k}$ perpendicular and parallel to the mean magnetic field, $\omega$ is the wave frequency. $I^A({\bf k})$ is the energy spectrum of the Alfv\'en modes and $I^M({\bf k})$ represents the energy spectrum of magnetosonic modes, which in our case at hand are the pseudo-Alfv\'en modes. In QLT, the resonance function $R_n=\pi\delta(k_{\parallel}v_{\parallel}-\omega\pm n\Omega)$. Now due to the perturbation of the orbit, it should be   
\bea
&&R_n(k_{\parallel}v_{\parallel}-\omega\pm n\Omega)\nonumber\\
&=&\Re\int_0^\infty dt e^{i(k_\|v_\|+n\Omega-\omega) t-\frac{1}{2}k_\|^2v_\bot^2t^2 \left(\frac{<\delta B_\parallel^2>}{B_0^2}\right)^\frac{1}{2}}\nonumber\\
&=&\frac{\sqrt{\pi}}{|k_\|\Delta v_\||}\exp\left[-\frac{(k_\|v \mu-\omega+n\Omega)^2}{k_\|^2\Delta v_\|^2}\right]\nonumber\\
&\simeq&\frac{\sqrt{\pi}}{|k_\||v_\bot M_A^{1/2}}\exp\left[-\frac{(k_\|v \mu-\omega+n\Omega)^2}{k_\|^2v^2(1-\mu^2)M_A}\right]
\label{resfunc}
\eea
where $M_A=\delta V/v_A=\delta B/B_0$ is the Alfv\'enic Mach number and $v_A$ is the Alfv\'en speed. We stress that Eqs.~(\ref{general},\ref{resfunc}) are generic, and applicable to both incompressible and compressible medium. 

For gyroresonance ($n=\pm 1,2,...$), apparently the result is similar to that from QLT for $\mu\gg \Delta \mu=\Delta v_\|/v$. In this limit, Eq.(\ref{general}) represents a sharp resonance and becomes equivalent to a $\delta$-function when put into Eq.(\ref{general}).  
In general, the result is different from that of QLT, especially at $\alpha\rightarrow 90^o$, the resonance peak happens at $k_{\|,res}\sim \Omega/\Delta v$ in contrast to the QLT result 
$k_{\|,res}\sim\Omega/v_\|\rightarrow \infty$\footnote{We shall
show below, that due to the anisotropy, the scattering coefficient $D_{\mu\mu}$ is still very small if the Alfv\'en and the pseudo-Alfv\'en modes are concerned.}.

On the other hand, the dispersion of the $v_\parallel$ means that CRs with a much wider range of pitch angle can be scattered by the pseudo-Alfv\'en modes through TTD 
($n=0$), which is marginally affected by the anisotropy and much more efficient than the gyroresonance. Below we shall consider both the cases for scattering in strong and weak turbulence. 

The nonlinear approach we use here is based on particle trapping by large scale magnetic perturbations (V\"olk 1973,1975). The difference is that we have a Gaussian profile (Eq.\ref{resfunc}) resonance and he adopted a Heaviside step function. Formally our approach also has a similarity to the second order quasilinear theory that Shalchi (2005a) proposed for slab modes although his approach is based on a different set of approximations.

\subsection{Strong MHD turbulence}

\begin{figure}
\plotone{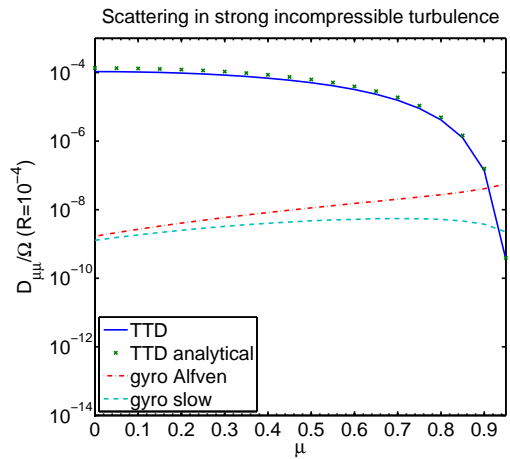}
\caption{Scattering of CRs with $R=r_L/L=10^{-4}$ in the strong incompressible turbulence ($M_A\simeq 1$). Solid line represents TTD and 'x' line is its analytical approximation (Eq.\ref{TTDstrong}). Dashdot refers to the gyroresonance with the Alfv\'enic turbulence while dash line is for the gyroresonance with the slow modes. }
\label{strongturb}
\end{figure}

In strong MHD turbulence, we assume
that the Alfv\'en and the pseudo-Alfv\'en modes follow the scaling obtained in Cho et al. (2002), which is consistent with the GS95 model:
\be
I^A({\bf k})=I^S({\bf k})=\frac{L^{-1/3}M_A^{4/3}}{6\pi}\exp\left(-\frac{L^{1/3}|k_\||}{M_A^{4/3}k_\bot^{2/3}}\right),
\ee
The pitch angle scattering arising from TTD with pseudo-Alfv\'en modes is given by Eq(\ref{general}) with $n=0$:

\bea
D^T_{\mu\mu}&=&\frac{v\sqrt{\pi}(1-\mu^2)}{6\pi LR^2}\int_{1}^{{\bf k}_{max}L}d^3x
\frac{x_\|x_\bot^{-10/3}M_A^{4/3}}{(x_\bot^2+x_\|^2)\Delta \mu}J_1^2(w)\nonumber\\
&&\exp\left[-\frac{x_\parallel}{x_\bot^{2/3} M_A^{4/3}}-\frac{(\mu-v_A/v)^2}{\Delta \mu^2 }\right],
\label{TTD}
\eea
where  $x=kL, x_\|=k_\|L$ are the normalized wavenumbers. ${\bf k}_{max}$ is the maximum wave vector of the turbulence, corresponding to the dissipation scale $k^{-1}_{max}$.
Since all scales contribute to TTD and the magnetic perturbation increases with scale, the interaction is dominated by the large scale moving mirrors ($k\sim 1/L$). Accordingly we can make an estimate of the above equation.  For CRs with small rigidities $R=v/(L\Omega)\ll 1$, $w=x_\bot R\sqrt{1-\mu^2}<1$, and $J_1(w)\sim w/2$. On the large scale, the anisotropy of the turbulence is small, and we approximate $(x_\bot^2+x_\|^2)$ in the above expression by $2 x_\bot^2$. Then we get from Eq.(\ref{TTD})

\bea
D^T_{\mu\mu}&\approx&\frac{\sqrt{\pi}M_A^{\frac{7}{2}}v}{16L}(1-\mu^2)^{\frac{3}{2}}\left[-E_1(qx_\|)-e^{-qx_\|}\right]_{1}^{x_{\|,max}}\nonumber\\
&&\exp\left[-\frac{(\mu-v_A/v)^2}{\Delta \mu^2}\right]
\label{TTDstrong}
\eea
where $E_1$ represents the function of exponential integral $E_1(x)=\int_1^\infty dt \exp(-xt)/t$ and $q=(x_{\bot,max}M_A^2)^{-2/3}$. As shown in Fig.\ref{strongturb}, the above expression provides a good approximation for Eq.(\ref{TTD}).

\begin{figure}
\plotone{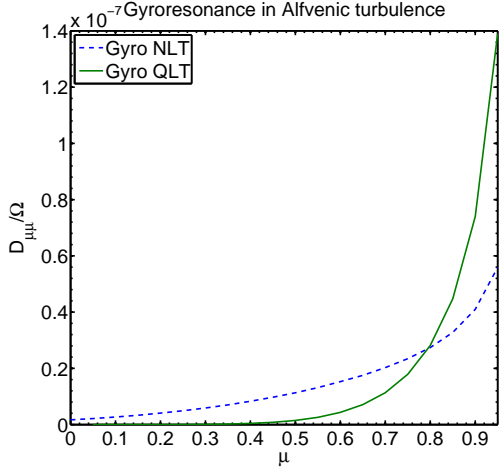}
\caption{Scattering of CRs with $R=r_L/L=10^{-4}$ by gyroresonance with the strong Alfv\'enic turbulence. The solid line represents the solution from nonlinear theory (Eq.\ref{gyro}), and the dashed line is the result from QLT (see YL02). }
\label{gyroqlt}
\end{figure}

{\it For gyroresonance}, the interaction is dominated by the first order harmonics because higher order harmonics operate on smaller scales where turbulence energy is decreasingly small according to the power law spectrum. The scattering with the Alfv\'en modes provides

\bea
D^G_{\mu\mu}&=&\frac{\sqrt{\pi} M_A^{\frac{4}{3}} v }{3\pi LR^2} (1-\mu^2)\int_{1}^{{\bf k}_{max}L}d^3x
\frac{x_\bot^{-\frac{10}{3}}}{x_\|\Delta \mu}\frac{J_1^2(w)}{w^2}\nonumber\\
&&\exp\left[-\frac{x_\parallel}{x_\bot^{2/3} M_A^{4/3} }-\frac{(\mu-\frac{1}{x_\|R})^2}{\Delta \mu^2}\right],
\label{gyro}
\eea
Eqs.(\ref{TTD}-\ref{gyro}) can be evaluated numerically. The results for the TTD scattering and gyroresonance are displayed in Fig.\ref{strongturb},\ref{gyroqlt}. We see that indeed taking into account the nonlinear perturbations, CRs of a wide range of pitch angles (including $90^o$) can be scattered through TTD interaction. At small pitch angles, TTD scattering is negligible and only gyroresonance operates. The result for gyroresonance is comparable to that from QLT (see Fig.\ref{gyroqlt}), showing again that gyroresonance is suppressed for the $k_\bot\gg k_\|$ turbulence. 

\subsection{Weak MHD turbulence}

In weak turbulence, smaller structure is developed only in the perpendicular direction. This entails an increase of $k_\bot$ while keeping $k_\|=L^{-1}$ unchanged. This type of cascade proceeds till ``critical balance'',  
$k_\|v_A\simeq k_\bot v_\bot$, is reached at the scale $l_{tr}$. The scaling of the weak turbulence is $E_k\propto k_\bot^{-2}$ (see Lazarian \& Vishniac 1999, appendix A there, Galtier et al. 2000). This renders 
\be
l_{tr}\sim LM_A^2
\label{ltr}
\ee 
and $v_{tr}\sim v_LM_A=v_AM_A^2$, 
where $M_A<1$. In other words, the weak turbulence only exists for a limited inertial range $L^{-1}<k_\bot<(LM_A^2)^{-1}$ 
(see also discussion in Lazarian 2006). 

If similar to the case of strong MHD turbulence, the pseudo-Alfv\'en modes in the weak turbulence follow the same scaling as the shear Alfv\'en modes, then Eq.(\ref{general}) provides
\bea
D^T_{\mu\mu}&=&\frac{v\sqrt{\pi}(1-\mu^2)}{2LR^2}\int_{1}^{M_A^{-2}}dx_\bot\frac{1}{(x_\bot^2+1)\Delta \mu}J_1^2(w) x_\bot^{-2}\nonumber\\
&&\exp\left[-\frac{(\mu-v_A/v)^2}{\Delta \mu^2}\right]
\label{TTDweak}
\eea 

\begin{figure}
\plotone{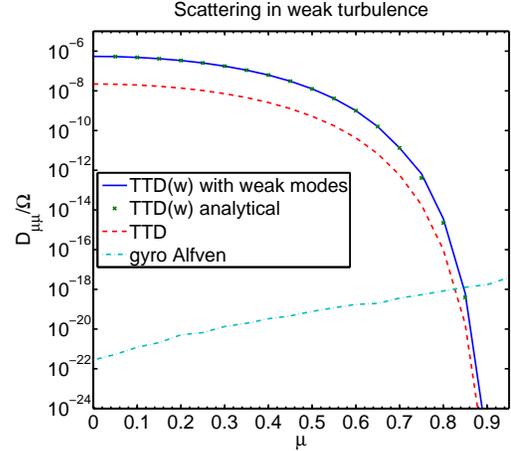}
\caption{Scattering coefficients in weak incompressible turbulence. Dashed line represents TTD (i.e., $n=0$) in the weak turbulence on large scales. 
Solid line and 'x' line (analytical approximation from Eq.\ref{TTDweaka}) refer to TTD with the weak pseudo Alfv\'en modes on large scales and dashed line  represents TTD with the strong turbulence on small scales. We see that gyroresonance (i.e., $n$ not equal to zero) which is denoted by the dashdot line is negligible because of the strong anisotropy. }
\label{weakalfv}
\end{figure}

In the case the rigidity R is much less than 1, the Bessel function $J_1^2(w)$ can be replaced by the first order approximation $w^2/4$. The above integration can be then evaluated analytically,

\bea
D^T_{\mu\mu}&=&\frac{v\sqrt{\pi}}{8L}(1-\mu^2)^{\frac{3}{2}}M_A^{\frac{3}{2}}[\tan^{-1}(M_A^{-2})-\pi/4]\nonumber\\
&&\exp\left[-\frac{(\mu-v_A/v)^2}{(1-\mu^2)M_A}\right]
\label{TTDweaka}
\eea
The result is compared with the numerical evaluation of Eq.(\ref{TTDweak}) in Fig.\ref{weakalfv}.
Fig.\ref{weakalfv} presents the scattering coefficients owing to various interactions in an incompressible medium with $M_A=0.1$. We account for TTD interactions with both the weak turbulence on large scales ($L\geq k^{-1} \geq l_{tr}$) and the strong turbulence on small scales ($k^{-1}<l_{tr}$). Here TTD is present for a smaller range of $\mu$ compared to the case of $M_A\simeq 1$ (Fig.\ref{strongturb}), as $\Delta \mu\sim M_A^{1/2}\ll\mu$. For the rest of the pitch angles, the turbulence can only scatter CRs through gyroresonance, which is inefficient because of the turbulence anisotropy. 

\section{CR scattering in compressible MHD turbulence}

\begin{figure}
\includegraphics[ width=0.8\columnwidth]{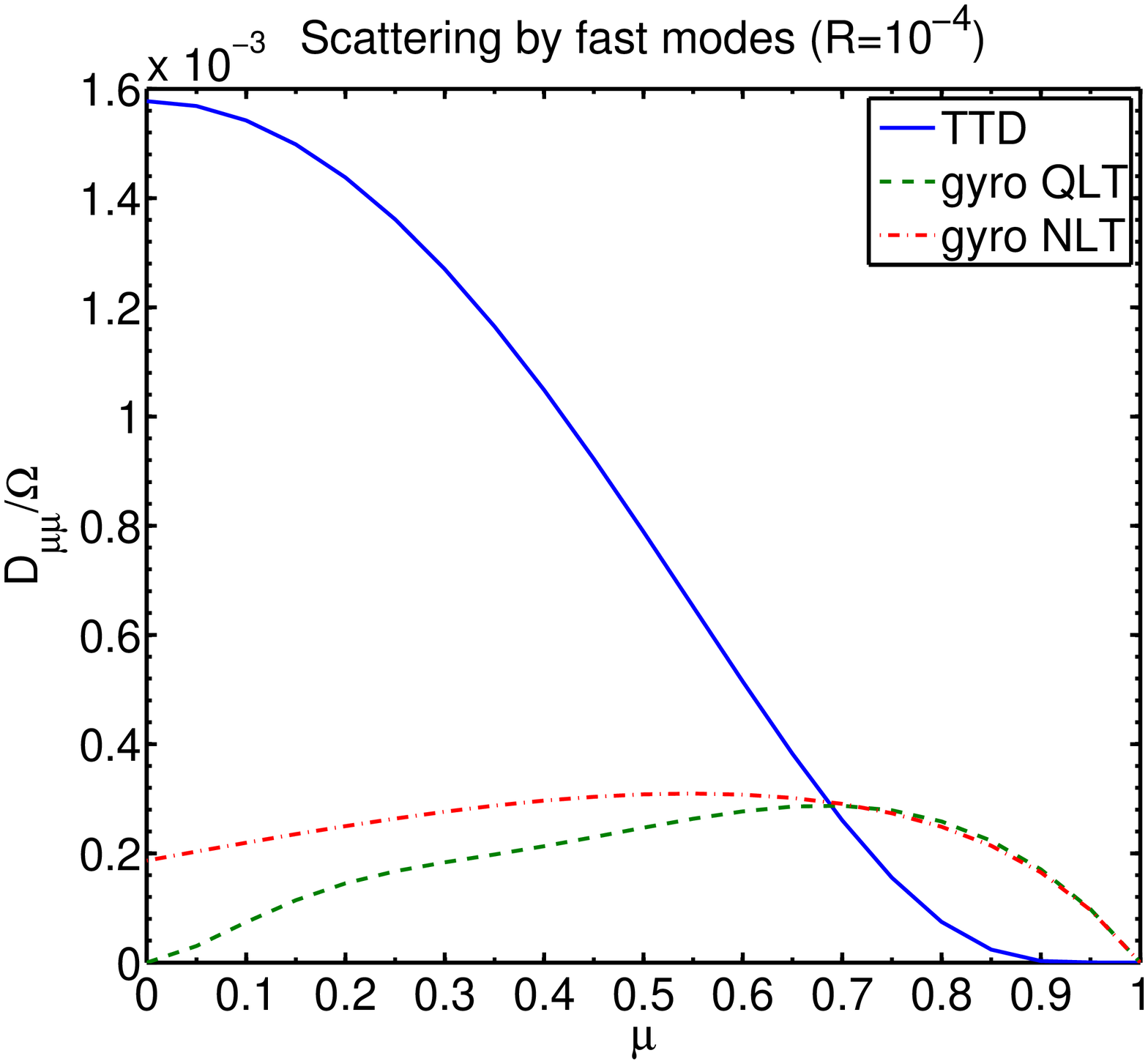}\vfil
\includegraphics[  width=0.8\columnwidth]{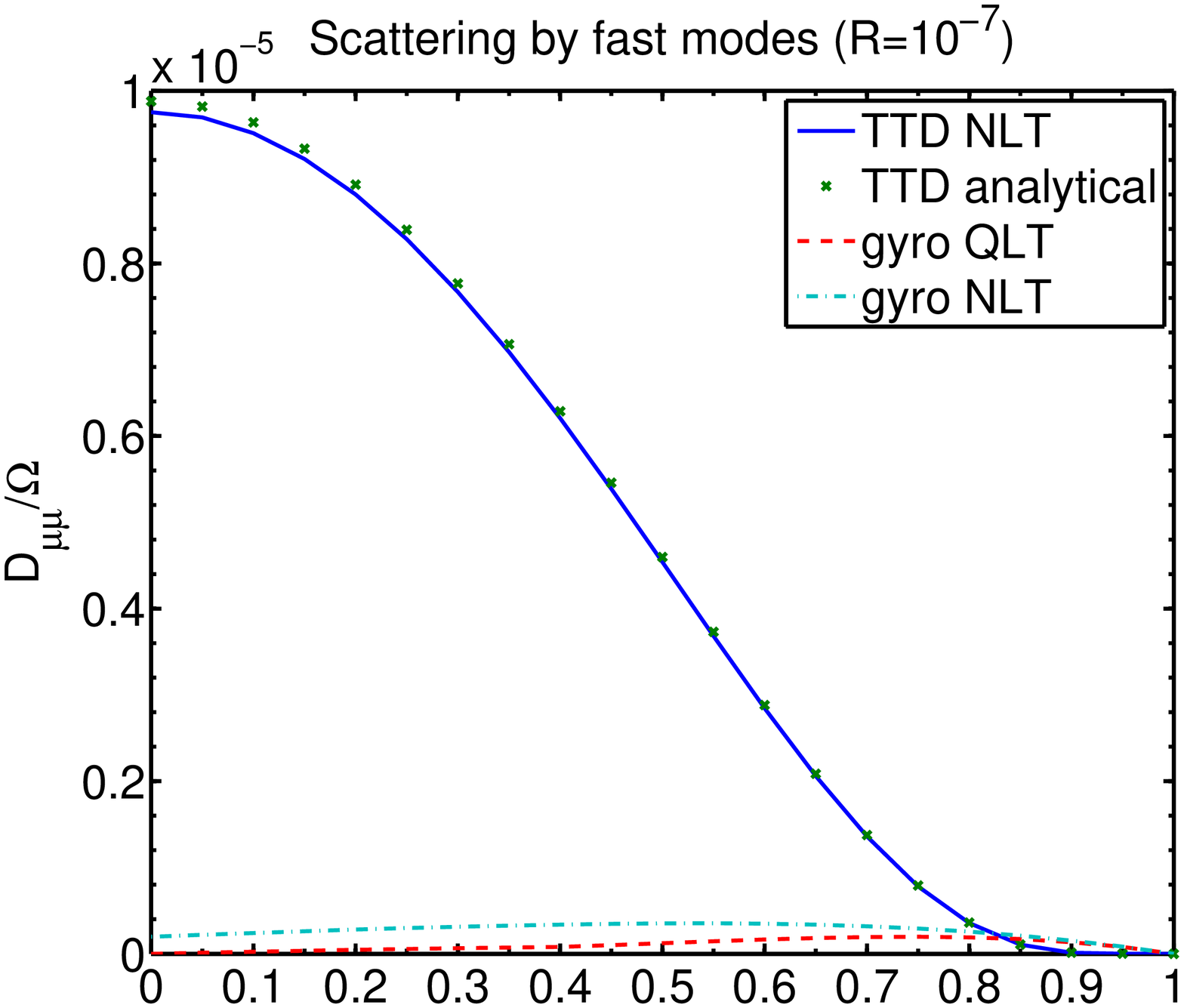}
\caption{Scattering of CRs in the fast mode turbulence. The solid line is the result for TTD based on NLT (Eq.\ref{fastTTD}) and the 'x' line represents its analytical approximation (Eq.\ref{lbttd}); the dashdot (NLT, Eq.\ref{fastgyro}) and dashed (QLT, Paper I) lines refer to the results of gyroresonance. {\it Upper}: scattering coefficient for CRs of $R=r_L/L=10^{-4}$; {\it Lower}: scattering coefficient for CRs of $R=r_L/L=10^{-7}$. This rigidity is enough small that Eq.(\ref{lbttd}) provides a good approximation.}
\label{fastcompr}
\end{figure}

In compressible turbulence, the pseudo-Alfv\'en modes become the slow modes, and additional type of perturbation, fast modes are present.
The latter were identified as the major scattering agent for the MHD turbulence that is injected at large scales (YL02,04). The papers were using QLT, however. Therefore it is necessary to check the validity of this conclusion using the modified resonance function given by Eq.(\ref{resfunc}). Therefore below we provide calculations for scattering induced by the fast modes. The scatterings by the Alfv\'en and the slow modes are similar to those by the Alfv\'en and the pseudo-Alfv\'en modes discussed in the previous section. The latter is the similarities of the former claimed on theoretical grounds (GS95) and confirmed using numerical simulations (CL03). 

For {\it gyroresonance}, according to Eq.(\ref{general},\ref{resfunc}), the pitch angle diffusion coefficient is given by
\bea
D^G_{\mu\mu}&=&\frac{v\sqrt{\pi}(1-\mu^2)}{2LR^2}\int_{1}^{k_{max}L}dx\int_0^1 d\xi
\frac{x^{-\frac{5}{2}}\xi}{\Delta \mu_\|}[J_1(w)']^2\nonumber\\
&&\exp\left[-\frac{(\mu-\frac{1}{x\xi R})^2}{\Delta \mu^2}\right],
\label{fastgyro}
\eea

where $\xi=\cos\theta$ is the cosine of wave pitch angle $\theta$.   

For {\it TTD}, 
\bea
D^T_{\mu\mu}&=&\frac{v\sqrt{\pi}(1-\mu^2)}{2LR^2}\int_{1}^{k_{max}L}dx\int_0^1 d\xi
\frac{x^{-5/2}\xi}{\Delta \mu_\|}J_1^2(w)\nonumber\\
&&\exp\left[-\frac{(\mu-v_A/v)^2}{\Delta \mu^2}\right],
\label{fastTTD}
\eea

For CRs with sufficient small rigidities $R<1/(k_{max}L)$, the Bessel function can be approximated by the first order 
asymptotics and we obtain: 
\bea
D^T_{\mu\mu}&=&\frac{\sqrt{\pi}}{4}v(1-\mu^2)^{\frac{3}{2}}\exp\left[-\frac{(\mu-v_A/v)^2}{\Delta \mu^2}\right]\nonumber\\
&&\int_0^1 d\xi \sqrt{\frac{k_{max}(\xi)}{L}}\xi(1-\xi^2),
\label{lbttd}
\eea
where $k_{max}(\xi)$ is the cut-off wave number at $\xi$.

The scattering by the fast modes is exhibited in Fig.\ref{fastcompr}, where we provided the numerical evaluations of Eqs(\ref{fastgyro},\ref{fastTTD}) as well as the analytical appoximation given by Eq.(\ref{lbttd}). We adopt viscous damping for the illustrative calculations (with the physical parameters in WIM, see table~1). Realistic calculation for interstellar medium will be given in the next section. As we see, the TTD interaction dominates for large pitch angles till $90^o$. Gyroresonance is important for small pitch angles. While the TTD interaction is expanded to a much wider range (including $90^o$) compared to QLT result (see, e.g, Fig.3 in Paper I), the gyroresonance in QLT and nonlinear theory (NLT) are comparable. This is because gyroresonance happens on a local scale $k_\|^{-1}\sim r_L$ unlike TTD. The influence of large scale trapping is thus limited. Especially for the range $\mu>\Delta \mu$, we see marginal difference. For large pitch angles, indeed there is a discrepancy between NLT and QLT. In fact, we see similar trend in the result of Shalchi (2005a), which has close relations with the nonlinear theory by V\"olk (1973, 1975). However, because of the dominance of TTD in this range, this difference does not count. All in all, nonlinear effect is important for CR scattering, particularly for the contribution from TTD interaction; for gyroresonance, nevertheless, one can use the quasi-linear approximation to calculate the scattering of CRs at small pitch angles.

Note that although the overall contribution from gyroresonance is smaller than that from TTD, gyroresonance plays an important role in confining the CRs at small pitch angles. Without sufficient scattering by gyroresonance (e.g., with the incompressible turbulence, see Fig.\ref{strongturb} ), the mean free path would be unrealistically large as TTD is inefficient for the scattering of CRs propagating at small pitch angles. Below in the next section we shall study the confinement of CRs in the Galaxy by the fast modes.

\section{Cosmic ray propagation in Galaxy}

The scattering by fast modes is influenced by the medium properties as the fast modes are subject to linear damping, e.g., Landau damping (see Ginzburg 1961).
In Paper I we showed that the CR scattering is different in different ISM phases, but could not calculate the mean free path as we faced the scattering at $90^o$ problem. Using the approach above we revisit the problem of the CR propagation in the selected phases of the ISM. In particular, we shall make quantitative predictions for the parallel mean free path of CRs in the galactic Halo and Warm Ionized medium (see Table~1 for a list of fiducial parameters appropriate for the idealized phases\footnote{The parameters of idealised interstellar phases are a subject of debate. Recently, even the entire concept of the phase being stable
 entities has been challenged (see Gazol et al. 2007 and ref. therein). Similarly as
 in Paper I, we were guided in choosing the numbers by our communications with Don
 Cox (2006, private communication). However, we accept that different parts
 of interstellar medium can exhibit variations of these parameters (see
 Wolfire et al. 2003 and ref. therein).}) assuming that the turbulence is injected at large scales.

\begin{figure}
\plotone{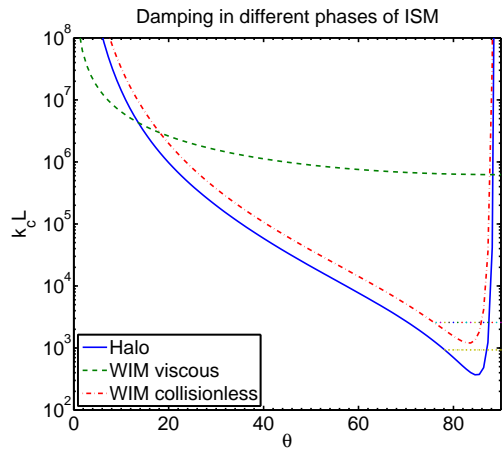}
\caption{The turbulence truncation scales in Galactic halo and warm ionized medium (WIM). The damping curves flattens around $90^o$ due to field line wandering (dotted lines, see Paper I, Lazarian, Vishniac \& Cho 2004); For WIM, both viscous and collisionless damping are applicable. }
\label{dampcurv}
\end{figure}

\begin{figure}
\plotone{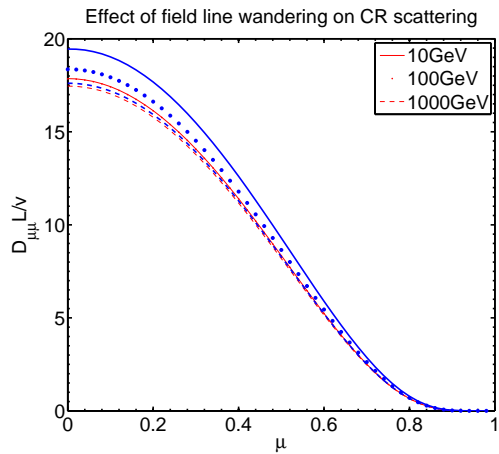}
\caption{Scattering efficiency is reduced for moderate energy ($\lesssim 1$TeV) CRs due to the average of the damping scale around $90^o$ caused by the field line wandering. The thick lines refer to the results without field line wandering while the thin lines represent the results taking into account field line wandering.}
\label{comprfw}
\end{figure}

\begin{table*}
\begin{tabular}{|c|c|c|c|c|c|c|c|c|}
\hline 
\hline 
ISM&T(K)&$C_S$(km/s)&n(cm$^{-3}$)&$l_{mfp}$(cm)&L(pc)&B($\mu$G)&$\beta$&damping
 \tabularnewline
\hline
Halo&$10^6$&91&$10^{-3}$&$4\times 10^{19}$&100&5&0.14&Collisionless\\
\hline
WIM& 8000&8.1&0.1&$6\times10^{12}$&50&6&0.077&Collisionless \& viscous
\tabularnewline
\hline
\hline
\end{tabular}
\caption{The fiducial parameters of idealized ISM phases and the relevant dampings. The dominant damping mechanism for the fast modes turbulence is given in the last line. WIM=warm ionized medium.}
\end{table*}

\subsection{Halo}

In Galactic halo (see Table~1), the Coulomb collisional mean free path is $\sim 10$pc, the plasma is thus in a collisionless regime. The cascading rate 
of the fast modes is (Cho \& Lazarian 2002).
\be
\tau_k^{-1}=(k/L)^{1/2}\delta V^2/V_{ph}
\label{tcasfast}
\ee

By equating it with the collisionless damping rate 
\bea
\Gamma_{c} & = & \frac{\sqrt{\pi\beta}\sin^{2}\theta}{2\cos\theta}kv_A\times \left[\sqrt{\frac{m_e}{m_i}}\exp\left(-\frac{m_e}{\beta m_i\cos^2\theta}\right)\right.\nonumber\\
& +& \left.5\exp\left(-\frac{1}{\beta\cos^{2}\theta}\right)\right],
\label{Ginz}
\eea
we obtain the turbulence truncation scale $k_c=k_{max}$:
\be
k_c L=\frac{4M_A^4m_i\cos^2\theta}{\pi m_e\beta\sin^4\theta}\exp\left(\frac{2m_e}{\beta m_i\cos^2\theta}\right).
\label{landauk}
\ee
where $\beta=P_{gas}/P_{mag}$.

The scale $k_c$ depends on the {\it wave pitch angle} $\theta$, which makes
the damping anisotropic. As the turbulence undergoes turbulent cascade or/and the waves propagate in a turbulent medium, the angle $\theta$ is changing.
As discussed in Paper I the field wandering defines the spread of angles. During one cascading time, the fast modes propagate a distance 
$v\tau_{cas} $ and see an angular deviation $\tan \delta \theta \simeq \sqrt{\tan^2\delta \theta_\parallel+\tan^2 \delta\theta_\perp}$, which is
\be
\tan \delta\theta \simeq  \sqrt{\frac{M_A^2\cos\theta}{27(kL)^{1/2}}+\left(\frac{M_A^2\sin^2\theta}{kL}\right)^{1/3}}
\label{dthetaB}
\ee
As evident, the damping scale given by Eq.(\ref{landauk}) varies considerably especially when $\theta\rightarrow 0$ and $\theta\rightarrow 90^o$. For the quasi-parallel modes, the randomization ($\propto (kL)^{-1/4}$) is negligible since the turbulence cascade continues to very small scales. On small scales, most energy of the fast modes is contained in these quasi-parallel modes (Paper I, Petrosian, Yan \& Lazrian 2006).

For the quasi-perpendicular modes, the damping rate (Eq.\ref{Ginz}) should be averaged over the range $90^o-\delta\theta \thicksim 90^o$. Equating Eq.(\ref{tcasfast}) and Eq.(\ref{Ginz}) averaged over $\delta\theta$, we get the averaged damping wave number (see Fig.\ref{dampcurv}). The field line wandering has a marginal effect on the gyroresonance, whose interaction with the quasi-perpendicular modes is negligible (Paper I). However, TTD scattering rates of moderate energy CRs ($<10$TeV) will be decreased owing to the increase of the damping around the $90^o$ (see Fig.\ref{comprfw}). For higher energy CRs, the influence of damping is marginal and so is that of field line wandering. 

We adopt QLT for calculating the gyroresonance (see Paper I). As we see from Fig.\ref{fastcompr}, the QLT result in the range $\mu>\Delta \mu$ provides a good approximation for our calculations using the non-linear approximation given by Eq.(\ref{fastgyro}). For CRs with sufficiently small rigidities, the resonant fast modes ($k_{res}\approx 1/(R\mu)$) are on small scales with a quasi-slab structure (see Fig.\ref{dampcurv}). For the scattering by these quasi-parallel modes, the analytical result that follows from QLT approximation (see Paper I) for the gyroresonance is\footnote{It can be shown that the QLT result follows from our more general result Eq.(\ref{fastgyro}) if we put $\Delta \mu \rightarrow 0$.} 

\begin{eqnarray}
\left[\begin{array}{c}
D^{G}_{\mu\mu}\\
D^{G}_{pp}\end{array}\right]&=&\frac{\pi v \mu^{0.5}(1-\mu^{2})}{4LR^{0.5}}\nonumber \\
&&\left[\begin{array}{c}
\frac{1}{7}[1+(R\mu)^2]^{-\frac{7}{4}}-(\tan^{2}\theta_c+1)^{-\frac{7}{4}}\\
\frac{m^{2}V_{A}^{2}}{3}\left\{[1+(R\mu)^2]^{-\frac{3}{4}}-(\tan^{2}\theta_c+1)^{-\frac{3}{4}}\right\}\end{array}\right]
\label{lbgyro}\end{eqnarray}
where $\tan\theta_c={k_{\perp,c}}/{k_{\parallel,res}}$. This justifies our use of the analytical approximation above.

\begin{figure}
\plotone{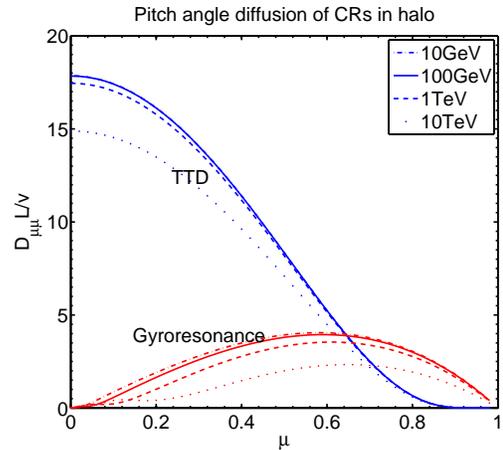}
\caption{Pitch angle diffusion coefficients in halo. Upper lines in the plots represent the contribution from TTD and lower lines are for gyroresonance.}
\label{Dmuhalo}
\end{figure}

\begin{figure}
\plotone{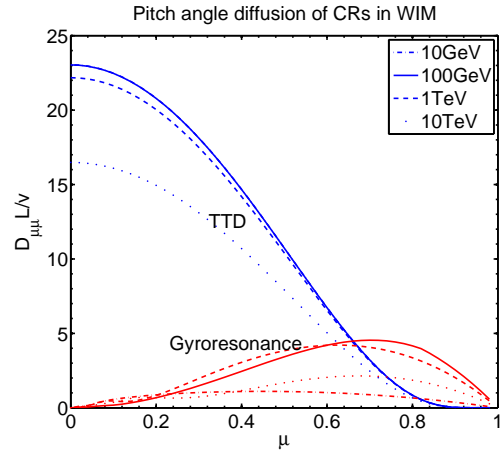}
\caption{Pitch angle diffusion coefficients in WIM. The notations are the same as in Fig.\ref{Dmuhalo}.}
\label{Dmuwim}
\end{figure}

\begin{figure}
\plotone{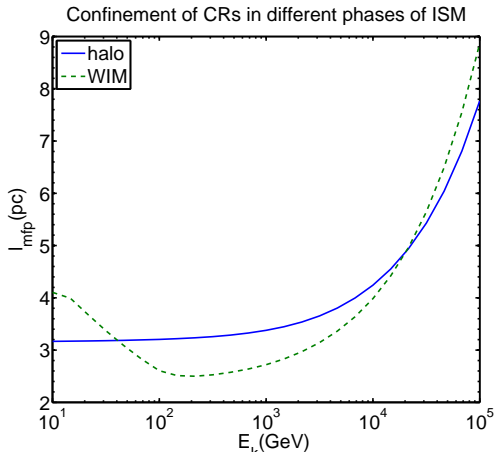}
\caption{ The mean free paths in two different phases of ISM: halo (solid line) and WIM (dashed line). At lower energies ($\lesssim100$GeV), the different dependence in WIM is owing to the viscous damping (see Fig.\ref{dampcurv} and text).}
\label{mfp}
\end{figure}

Once we the know the functional form of the $D_{\mu\mu}$, we can obtain the corresponding mean free path (Earl 1974):
\be
\lambda_\|/L=\frac{3}{4}\int^1_0 d\mu \frac{v(1-\mu^2)^2}{(D^T_{\mu\mu}+D^G_{\mu\mu})L},
\ee
where $D^T_{\mu\mu}$ can be obtained by Eqs.(\ref{fastTTD}-\ref{lbttd}, \ref{landauk}). Insert $D^T_{\mu\mu}$ and QLT result for $D^G_{\mu\mu}$ into the above expression, we get the mean free path of CRs in halo.

The mean free path is sensitive to the scattering by gyroresonance at small pitch angles, due to the influence of damping on the fast modes on small scales. Fig.\ref{Dmuhalo} shows the pitch angle diffusion of CRs with different energies due to the TTD and gyroresonance.
 
The weak dependence of the mean free path (see Fig.\ref{mfp}) of the moderate energy (e.g$<1$TeV) CRs in halo results from the  fact that gyroresonance changes marginally with the CR energy (see Fig.\ref{Dmuhalo}). Gyroresonance happens on small scales where the fast modes develop a quasi-slab structure because of the damping (see Fig.\ref{dampcurv}). In the case of halo, the critical $\theta_c$ changes more slowly compared to the case in WIM (see Fig.\ref{mfp}), the scattering by gyroresonance is thus marginally changing with the energy (see Eq.\ref{lbgyro} and Fig.\ref{Dmuhalo}). For higher energy CRs with larger gyroscales, the influence of damping is small, and thus the CR mean free path begins increasing with energy.

\subsection{Warm Ionized Medium}

In warm ionized medium, the Coulomb collisional mean free path is $l_{mfp}=6\times 10^{12}$cm and the plasma $\beta\simeq0.11$. Suppose that the turbulence energy is injected from large scale, then the compressible turbulence is subjected to the viscous damping besides the collisionless damping. 
By equating the viscous damping rate with the cascading rate (Eq.\ref{tcasfast}), we obtain the following truncation scale, 
\bea
k_{c}L=x_c\left\{\begin{array}{rl}(1-\xi^2)^{-\frac{2}{3}} & \beta\ll 1\\
(1-3\xi^2)^{-\frac{4}{3}} & \beta\gg 1\end{array}\right.
\eea
where $x_c=\left[\frac{6\rho\delta V^2L}{\eta_0V_A}\right]^{\frac{2}{3}}$,
 $\eta_0$ is the longitudinal viscosity. In the low $\beta$ regime, the motions are primarily perpendicular to the magnetic field so that $\partial v_{x}/\partial x=\dot{n}/n\sim\dot{B}/B$. The longitudinal viscosity enters here as the result of distortion of the Maxwellian distribution (see Braginskii 1965). The transverse energy of the ions increases during compression because of the conservation of adiabatic invariant $v_{\perp}^{2}/B$. If the rate of compression is faster than that of collisions, the ion distribution in the momentum space is bound to be distorted from the Maxwellian isotropic sphere to an oblate spheroid with the long axis perpendicular to the magnetic field. As a result, the transverse pressure gets greater than the longitudinal pressure, resulting in a stress $\sim\eta_{0}\partial v_{x}/\partial x$.
The restoration of the equilibrium increases the entropy and causes the dissipation of energy.

The viscous damping scale is compared to collisionless cutoff scale (Eq.\ref{landauk}) in Fig.\ref{dampcurv}. 
As shown there, both viscous damping and collisionless damping are important in WIM. Viscous damping is dominant for small $\theta$ and
 collisionless damping takes over for large $\theta$ except for $\theta=90^o$.
This is because collisionless damping increases with $\theta$ much faster than the viscous damping. For sufficiently small wave pitch angles, the viscous damping is too small to prevent the fast modes to cascade down to scales smaller than the mean free path $l_{mfp}$. Because of the similar quasi-slab structure on small scales, 
Eq.(\ref{lbgyro}) can be also applied in WIM. The results are illustrated in Fig.\ref{Dmuwim}. Compared to the case in halo, we see that the qualitative difference stands in the gyroresonance. This is because gyroresonance is sensitive to the quasi-slab modes for which dampings differ in halo and WIM.  

The mean free paths of CRs are given in Fig.\ref{mfp}. We see the mean free path first decreases with the energy till 100GeV. This is because of the influence of viscous damping on gyroresonance (see Fig.\ref{dampcurv}). 
The lower the energy of the CRs is, the narrower is the wave vector cone of the available fast modes and thus the less efficient is the gyroresonance according to Eq.(\ref{lbgyro}) (see Fig.\ref{Dmuwim}). For higher energy CRs, of which rigidities are larger than $R\gtrsim(3\times 10^6)^{-1}$, the maximum wave number of the resonant wave modes are determined by collisionless damping (see Fig.\ref{dampcurv}). As a result, the mean free path grows with energy similar to the case in halo where collisionless damping is dominant.

\subsection{Other phases}

In hot ionized medium (HIM), the plasma is also in collisionless regime, but the density is higher and the plasma beta is larger than 1. The damping by protons thus becomes substantial especially at small pitch angles. The damping truncates the turbulence at much larger scales than the gyroscales of the CRs of the energy range we consider. No gyroresonance can happen and some other mechanisms are necessary to prevent CRs streaming freely along the field. The turbulence injected from small scales might play an important role (see \S5).

In partially ionized gas one should take into account an additional damping that arises from ion-neutral collisions (see Kulsrud \& Pearce 1969, Lithwick \& Goldreich 2001, Lazarian, Vishniac \& Cho 2004). In the latter work a viscosity-damped regime of turbulence was predicted at scales less the scale $k_{c, amb}^{-1}$ at which the ordinary magnetic turbulence is damped by ionic viscosity. The corresponding numerical work, e.g., Cho, Lazarian \& Vishniac (2003) testifies that for the viscosity-damped regime the parallel scale stays equal to the scale of the ambipolar damping, 
i.e., $k_{\|}=k_{c, amb}$, while $k_{\bot}$ increases. In that respect, the scattering by such magnetic fluctuations is analogous to the scattering induced by the weak turbulence (see \S 2.3). The difference stems from the spectrum
of $k_{\bot}$ is shallower than the spectrum of the weak turbulence. The predicted values of the spectrum for the viscosity-damped turbulence $E(k_\bot)\sim
k_\bot^{-1}$ (Lazarian et al. 2004) are in rough agreement with simulations. More detailed studies of scattering in partially ionized gas will be provided elsewhere.

\section{Perpendicular transport}

While in the earlier sections we dealt entirely with the diffusion parallel to the magnetic field, in this section we deal with the diffusion perpendicular to the {\it mean} magnetic field. The assumption that
 CRs follow the magnetic field averaged over their Larmor radius is pretty accurate in most situations, e.g., for Galactic CRs. 

Compound diffusion happens when particles are restricted to the magnetic field lines and perpendicular transport is solely due to the random walk of field line wandering (see K\'ota \& Jokipii 2000). 
In the three-dimensional turbulence, field lines are diverging away due to shearing by the Alfv\'en modes (see Lazarian \& Vishniac 1999, Narayan \& Medvedev 2002, Lazarian 2006, 2007).
 Since the Larmor radii of CRs are much larger than the minimum scale of eddies $l_{\bot, min}$, field lines within the CR Larmor orbit are effectively diverging away owing to shear by the Alfv\'enic turbulence.
The cross-field transport thus results from the deviations of field lines at small scales, as well as field line random walk at large scale ($>{\rm min}[L/M^3_A,L]$).

Most recently the diffusion in magnetic fields was considered for thermal particles in Lazarian (2006, 2007). In what follows we modify the results of these studies for the case of CRs.

For perpendicular diffusion, the important issue is the reference of frame. We emphasize that we consider the diffusion perpendicular to the {\emph mean} field direction in the global reference of frame.

\subsection{Perpendicular diffusion on large scale}

{\it High $M_A$ turbulence}:\\
High $M_A$ turbulence corresponds to the field that is easily bended by the hydrodynamic motions at the injection scale as the hydro energy at the injection scale is much larger than the magnetic energy, i.e.
$\rho V_L^2\gg B^2$. The turbulence in clusters of galaxies is the high $M_A$ turbulence. In this case the magnetic field becomes dynamically important on a much smaller scale, i.e., the scale $l_A=L/M_A^3$ (see Lazarian 2006). If $\lambda_\|\gg l_A$, the CR diffusion is controlled by the straightness of the field lines, and
\be 
D_\bot=D_{\|}\approx 1/3l_Av,~~~M_A>1,~~~\lambda_{\|}>l_A.
\label{dbb}
\ee
The diffusion is isotropic if scales larger than $l_A$ are concerned.

In the opposite limit $\lambda_{\|}<l_A$, the stiffness of B field is negligible. The CR diffusion is insensitive to the topology of the magnetic field. In the global reference of frame, there is no distinction between the perpendicular and parallel direction. Naturally, a result for isotropic turbulence, namely,
\be
D_{\bot}= D_{\|}\sim 1/3 \lambda_{\|} v,
\ee
holds.

{\it Low $M_A$ turbulence}:\\
For strong magnetic field, i.e., the field that cannot be easily bended at the turbulence injection scale, individual magnetic field lines are aligned with the mean magnetic field. The diffusion in this case is anisotropic.
As we mentioned earlier, if the turbulence is injected at scale $L$ it stays weak for the scales larger than $l_{tr}$ given by Eq.~(\ref{ltr}) and it is strong at smaller scales. 

Consider first the case of the CR parallel mean free path larger than the injection scale of the turbulence, i.e., $\lambda_\|>L$. The perturbations of the field are uncorrelated over scales larger than $L M_A^2$ in the direction perpendicular to the mean magnetic field. Indeed, this perpendicular distance corresponds
to the particle moving parallel distance of the order $L$, which is the scale
of the energy injection, which in a simplified picture of turbulence\footnote{For the sake of simplicity we disregard the effects of the inverse cascade that
can increase the correlation scale of magnetic perturbations.} is the
maximal scale over which the magnetic perturbations are correlated.
 
In this situation the random walk steps in perpendicular direction are of $l_{tr}$ length. Thus, to diffuse over a distance R with random walk of $l_{tr}$ one requires $(R/l_{tr})^2$ steps.
The time of the individual step is $L/v_\|$, then 
\be
D_\perp=\frac{R^2}{\delta t}=\frac{R^2}{(R/l_{tr})^2L/v_\|}\approx 1/3Lv M_A^4, ~~~M_A<1,~~~ \lambda_\|>L. 
\ee
This is similar to the case discussed in the FLRW model (Jokipii 1966). However, we obtain the dependence of $M_A^4$ instead of their $M_A^2$ scaling. This difference is not crucial for the environment like Milky Way or solar wind, for which $M_A\sim 1$, but may be important for other environments where strong slightly perturbed magnetic field is present, e.g., solar corona.

What would be the CR diffusion perpendicular to the mean magnetic field in the opposite case of $\lambda_\|<L$?  The time of the individual step is $L^2/D_{\parallel}$. Therefore the perpendicular diffusion coefficient is
\be
D_{\bot}=\frac{R^2}{\delta t}\approx \frac{R^2}{(R/l_{tr})^2 L^2/D_{\parallel}}=D_{\|}M_A^4,~~~M_A<1,~~~\lambda_\|<L ,
\label{diffx}
\ee
 which coincides with the result obtained for the diffusion of electrons in magnetized plasma (Lazarian 2006). The turbulence in the interplanetary medium is in this regime with $M_A\lesssim 1$. From the Eq.(\ref{diffx}), we obtain a constant ratio of $\lambda_\bot/\lambda_\|=D_\bot/D_\|=M_A^4$, consistent with the Palmer consensus (Palmer 1982).

We mention parenthetically, that our arguments above can be repeated for any random walk process
in the perpendicular direction with a step $\delta x$. If we
can write $D_{\perp}=(\delta x/\delta z )^2 D_{\parallel}$, we, naturally, recover the result in Eq.~(\ref{diffx}).

\subsection{Perpendicular diffusion on small scales}

The diffusion of CR on the scales $\ll L$ may be different. We consider particular examples below. 

{\it High $M_A$ turbulence}:\\
Consider the diffusion on scales that are
On scales $\lambda_{\|}<k_\parallel^{-1}<l_A$, i.e., on scales at which CR are
in diffusive regime, but the magnetic fields are strong enough to influence
turbulent motions, 
the mean deviation of a field in a distance $k_\|^{-1}=\delta z $ is given by (Lazarian \& Vishniac 1999),
\be
<(\delta x)^2>^{1/2}=\frac{([\delta z] M_A)^{3/2}}{3^{3/2}L^{1/2}},~~~M_A>1
\label{highmwand}
\ee
Thus, for scales much less than $L$
\be
D_\bot\approx
\left(\frac{\delta x}{\delta z}\right)^2 D_{\|}\sim \frac{[\delta z] M_A^3}{3^3L} D_{\|}\sim D_\| (k_\|l_A)^{-1}, ~~~M_A>1, 
\label{dw1} 
\ee
which for a limiting case $k_{\|}\sim l_A$ gets the result consistent with 
Eq.~(\ref{dbb}). 

{\it Low $M_A$ turbulence}:\\
On scales larger than $l_{tr}$, the turbulence is weak (see \S 3.2). The mean
 deviation of a field in a distance $\delta z$ is given by Lazarian (2006):
\be
<(\delta x)^2>^{1/2}=\frac{[\delta z]^{3/2}}{3^{3/2}L^{1/2}}M_A^2,~~~M_A<1.
\label{lowmwand}
\ee

For the scales $L>k_\parallel^{-1}=\delta z>\lambda_{\|}$
we combine Eq.~(\ref{lowmwand}) with
\be
\delta z=\sqrt{D_\| \delta t}
\label{zz}
\ee
 and get for scales much less than $L$
\be
D_\bot\approx
\frac{\delta x^2}{\delta t}=\frac{D_\|\delta z}{3^3L}M_A^4\sim D_\|(k_\|L)^{-1}
M_A^4, 
\label{dw2}
\ee
which for a limiting case of $k_{\|}\sim L^{-1}$ coincides up to a factor
with the Eq.~(\ref{diffx}).

Eqs.~(\ref{dw1}) and (\ref{dw2}) certify that the perpendicular diffusion at scales much less than the injection scale accelerates as z grows. The reason is that there is no random walk on small scales up to the injection scale of the strong MHD turbulence ($l_{tr}$ for $M_A < 1$  and $l_A$ for $M_A>1$). The diffusion on 
scales less than the turbulent injection scale is important for describing 
propagation and acceleration of CR in supernovae shells, clusters of galaxies
etc.    

\subsection{Subdiffusion}
The diffusion coefficient in Eq.~(\ref{diffx}), i.e., $D_{\|}M_A^4$, means that the transport perpendicular to the dynamically strong magnetic field is a diffusion, rather than subdiffusion, as it was stated in a number of recent papers. Let us clarify this point by obtaining the necessary conditions for the subdiffusion to take place.

In the papers discussing compound diffusion 
(see K\'ota \& Jokipii 2000, Webb et al. 2006), 
$(\delta x)^2/\delta z=D_{spat}$ is a spatial diffusion constant coefficient. If we adopt this, we shall indeed get from Eq.~(\ref{zz}) the perpendicular diffusion coefficient
\be 
D_{\perp}=\left(\frac{\delta x}{\delta z}\right)^2D_\|=D_{spat}D_\|/\delta z=D_{spat} D_{\parallel}^{1/2} (\delta t)^{-1/2}
\label{subdiffusion}
\ee
Therefore the perpendicular transposition will be
$x^2=D_{\perp} \delta t= D_{spat}D_{\parallel}^{1/2} (\delta t)^{1/2}$ in accordance with the findings in the aforementioned papers.

The major implicit assumption in the reasoning above is that the particles trace back their trajectories in x direction on the scale $\delta z$. If this is not true, the introduction of the diffusion coefficient $D_{spat}$ does not make sense.

When is it possible to talk about tracing particle trajectories back? In the case of random motions at a single scale {\it only}, the distance over which the particle trajectories get uncorrelated is given by the Rechester \& Rosenbluth (1978) model. On scales larger than the Rechester \& Rosenbluth scale $k_\|^{-1}>L_{RR}$, the separation between field lines $\delta x$ grows monochromatically with the distance $\delta z$ along the magnetic field, no retracing can happen in this case¡­  Assuming that the damping scale of the turbulence is larger than the CR Larmor radius, this model, when generalized to the anisotropic turbulence provides (Narayan \& Medvedev 2001, Lazarian 2006)
\be
L_{RR}=l_{\|, min}\ln(l_{\bot, min}/r_L)
\ee
where $l_{\|, min}$ is the parallel scale of the cut-off of turbulent motions, 
$l_{\bot, min}$ is the corresponding perpendicular scale, $r_L$ is the CR Larmor radius. The assumption of $r_L<l_{\bot, min}$ can be valid, for instance, for the Alfv\'enic motions in partially ionized gas.
However, it is easy to see that, even in this case, the corresponding scale is rather small and therefore subdiffusion is not applicable for the transport of particles in the Alfv\'enic turbulence over scales $\gg l_{\|,min}$.

If  $r_L>l_{\bot, min}$, as it is a usual case for the Alfv\'en motions in the phase of ISM with the ionization larger than $\approx 93\%$, where the
Alfv\'enic motions go to the thermal particle gyroradius 
(see the estimates in Lithwick \& Goldreich 2001, Lazarian et al. 2004), the subdiffusion of CR is not an applicable concept for the Alfv\'enic turbulence.  
This does not preclude subdiffusion from taking place in particular models of magnetic perturbations, e.g., in the slab model considered in Shalchi (2005b), but we believe in the omnipresence of the Alfv\'enic turbulence in interstellar gas (see Armstrong et al. 1995).

\section{Discussions}

The present paper extends our study in Paper I. As in Paper I we mostly deal 
with the magnetic perturbations that are part of the
large scale turbulent cascade, which is consistent with the 
Big Power Law in the sky observed via radio-scattering and scintillation 
technique (Armstrong et al. 1995). In both papers we use the description of 
the MHD turbulence that follows from numerical simulations. 

 In Paper I we have the CR scattering calculated in the selected interstellar environments making use of Quasi-Linear Theory (QLT). Because of the limitations of the QLT, we could not provide calculations of the mean free path in Paper I, which limited the utility of the study. In this paper we extended the non-linear approach suggested in V\"olk (1975) to treat the scattering, which allows us to calculate the mean free paths that arise from CR interactions with the fast modes. In doing so, similar to Paper I, we take into account damping of the fast modes in the presence of the field wondering induced by the Alfv\'enic modes. 

Our results show that in WIM and halo of our Galaxy, confinement of bulk CRs are mostly due to the compressible modes. We obtain CR mean free paths about a few parsec, consistent with what observations indicate. The major difference with earlier picture is the dependencies of CR transport parameters on the medium properties. The dependence appears as a result of damping of the fast modes. For low energy CRs ($\lesssim 100$GeV), if dominated by viscous damping, the mean free path of CRs would decrease with energy; with collisionless damping, however, CRs' mean free path stays almost a constant. Field line wandering in general increases the damping of the fast modes and reduces the scattering efficiency of CRs. For higher energy CRs, the influence of damping is limited, and their mean free path increases with energy. 

The dependencies on the turbulence damping and therefore the phase properties should have various implications from ratio of secondary to primary elements, diffuse Galactic $\gamma$ ray emission, to the CMB synchrotron foreground. With precise measurements, the understanding of CMB is now constrained by our understanding of the foreground. The variation of CR index over the Galaxy may paralyze the synchrotron templates. Such variations can be addressed on the basis of the more elaborate CR propagation theory.

The importance of this study goes beyond the interstellar medium. For instance, Brunetti \& Lazarian (2007), treated acceleration of CRs for plasma in clusters of galaxies appealing to the fast modes, which is the approach to CRs similar to that in Paper I.  
We believe that the non-linear treatment may be useful for such cases as well.
In addition, stochastic acceleration by the MHD turbulence is a promising mechanism for generating high energy particles during solar flares (see, e.g., Petrosian \& Liu 2004, and references therein). An application to the acceleration of CRs in solar flares will be given in Yan, Lazarian \& Petrosian (2007, in preparation).

In our treatment we attempted to use the scalings that (a) are consistent with numerical calculations and (b) whose amplitudes we can estimate with a sufficient degree of precision. Therefore our present study does not deal with scattering of CRs by the fast modes on the scales $l>LM_A^2$, $M_A<1$, i.e., on the scales where the Alfv\'enic turbulence in the weak regime. It was suggested by Chandran (2005) that the weak fast modes at small pitch angles tend to steepen due to the coupling with the Alfv\'en modes. When the resulting scaling of the fast modes becomes clearer, our approach will be applicable to them.

We have not quantitatively dealt in the present paper with the case of the slab Alfv\'en modes created by instabilities\footnote{Streaming instability (see
Cesarsky 1980) is an example of such instability. However, 
the instability is suppressed by both ion-neutral damping (Kulsrud \& Pierce 1969) and the ambient turbulence (YL02, Farmer \& Goldreich 2004, Paper I, Lazarian \& Beresnyak 2006). Another example is the gyroresonance instability
discussed in the context of CRs in Lazarian \& Beresnyak (2006).}. 
The CR scattering by the perturbations created by those modes may dominate over the gyroresonance with the fast modes, especially for 
CRs of low energies, i.e., whose gyroresonance with the fast modes is inefficient due to the fast modes damping (see estimates in Lazarian \& Beresnyak
2006). Progress in quantitative description of the non-linear stages of the instabilities that can create slab modes should enable comprehensive models that include both the fast modes and the slab modes.

In addition, we addressed the issue of perpendicular diffusion, the issue that we have not dealt with in Paper I. We found, that similar to the case of thermal diffusion discussed in Lazarian (2006), the diffusion of CRs depends on the Alfv\'enic Mach number $M_A$. We found that the suppression of the perpendicular diffusion compared to the parallel one scales as $M_A^4$ for $M_A<1$. Approaching the issue of subdiffusion, we found that it is negligible for CRs in the Alfv\'enic turbulence.

\section{Summary}

Our result can be briefly summarized as follows
\begin{itemize}
\item Treatment of the scattering in both strong and weak MHD turbulence has been generalized to account for perturbations of the particle orbits. We found
that the non-linear treatment is essential for calculating mean free paths of
CRs.

\item Our calculations of scattering rates performed for different modes of MHD turbulence, assuming that the turbulence is injected at large scales, confirm the dominance of the fast modes for scattering of the bulk CRs in WIM and Galactic halo.

\item We obtained the relation between the CR diffusion coefficient parallel
to the magnetic field and the CR diffusion coefficient perpendicular to the magnetic
field. We show that CR transport perpendicular to the magnetic field depends strongly on the Alfv\'en Mach number of the turbulence. 

\end{itemize}

\begin{acknowledgments}
We are grateful to the anonymous referee for his/her valueble comments and suggestions. We thank Prof. Thompson for fruitful discussions. 
HY is supported by CITA and the National Science and Engineering Research Council of Canada. AL acknowledges the NASA grant X5166204101, the NSF grant ATM-0648699, as well as the NSF Center for Magnetic Self-Organization in Laboratory and Astrophysical Plasmas. \end{acknowledgments}

\end{document}